# Prediction Markets, Mechanism Design, and Cooperative Game Theory

**Vincent Conitzer**
Dept. of Computer Science
Duke University
Durham, NC 27708, USA

## Abstract

Prediction markets are designed to elicit information from multiple agents in order to predict (obtain probabilities for) future events. A good prediction market incentivizes agents to reveal their information truthfully; such incentive compatibility considerations are commonly studied in mechanism design. While this relation between prediction markets and mechanism design is well understood at a high level, the models used in prediction markets tend to be somewhat different from those used in mechanism design. This paper considers a model for prediction markets that fits more straightforwardly into the mechanism design framework. We consider a number of mechanisms within this model, all based on proper scoring rules. We discuss basic properties of these mechanisms, such as incentive compatibility. We also draw connections between some of these mechanisms and cooperative game theory. Finally, we speculate how one might build a practical prediction market based on some of these ideas.

## 1　Introduction

A *prediction market* [16] is a market created for the purpose of obtaining a subjective probability distribution, based on the information of multiple agents. To predict whether a particular event (say, the Democratic candidate winning the election) will happen, a common approach is to create a security that will pay out some predetermined amount (say, $1) if the event happens, and let agents trade this security until a stable price emerges; the price can then (arguably) be interpreted as the consensus probability that the event will happen. However, there are also other designs for prediction markets. Examples include dynamic parimutuel markets [14], as well as market scoring rules [10] (we will discuss the latter in more detail). A good prediction market ensures that agents are rewarded for contributing useful and accurate information.

To analyze prediction markets, it is often assumed that agents will act myopically. In a market based on securities, this corresponds to the assumption that an agent will buy if the price is below her current subjective probability (which takes what happened in the market so far into account), and will sell if the price is above her current subjective probability. Even under this assumption, it is theoretically possible that the market converges to a price that does not reflect the full combined information of all the agents. For example, suppose that agent 1 observes $x \in \{0, 1\}$, and agent 2 observes $y \in \{0, 1\}$. Let $z = 1$ if $x = y$, and $z = 0$ otherwise; and consider a prediction market that attempts to predict whether $z = 1$. In principle, the agents collectively have enough information to predict $z$ perfectly. However, if we assume $x$ and $y$ are drawn independently according to a uniform distribution, then each agent's subjective probability that $z = 1$ is $0.5$ regardless of the information she receives, so the price will remain stuck at $0.5$. (A similar example is given in [6].) Of course, this is a knife-edge example. If we modify the distribution so that $P(y = 1) = 0.51$, then, given an initial price of $0.5$, if agent 1 observes $x = 1$, she will start buying and drive the price up; whereas if she observes $x = 0$, she will start selling and drive the price down. From this behavior, agent 2 can infer what agent 1 observed, and as a result knows $z$, so that the market price will correctly converge to $0$ or $1$.

A difficulty is presented by the fact that strategic agents will not always behave myopically: they may try to manipulate the beliefs of the other agents, and thereby, the market price. Considering the modified example again, suppose that agent 1 changes her strategy to do the *opposite* of what she did before. That is, she will start *selling* if $x = 1$, and *buying* if $x = 0$. If agent 2 is not aware of agent 1's strategic behavior, he will be misled into drawing the wrong conclusion about $z$, and drive the price to the exact opposite of what it should be. This leaves agent 1 in an advantageous position: if the price has been driven to $0$, she can cheaply buy securities that in reality are worth $1$; whereas if the price has been driven to 1, she can sell securities that are worth 0 at a price of 1. A similar example is given in [13].

We may ask ourselves whether we can avoid these difficulties by designing the prediction market appropriately. The creation of markets that lead to good results in the face of strategic behavior is a topic that falls under *mechanism design*. A standard approach in mechanism design is to create *direct-revelation* mechanisms, in which agents directly report all their private information, that are *incentive compatible*, meaning that agents have no incentive to misreport their private information.

The high-level idea that concepts from mechanism design should be applicable to prediction markets is certainly not novel. The relation is clear: in both cases, there are multiple agents with private information, and the goal is to incentivize them to reveal the (relevant) information. For example, [11, 12] also study some versions of incentive compatibility and other mechanism design concepts in the context of prediction markets. Market scoring rules incentivize agents to report probabilities truthfully, if they are myopic [13, 2].

However, it appears that at this point, there is still somewhat of a gap between the theories of prediction markets and mechanism design. This becomes especially apparent when one considers, for contrast, how smoothly auction theory integrates with mechanism design. The goal of this paper is to reduce (or perhaps eliminate) the gap, by laying out a framework for prediction markets that fits better with the theory of mechanism design. In particular, we will consider a type of direct-revelation prediction market, in which agents report their full information directly (rather than trading securities, reporting probabilities, *etc.*). Like market scoring rules, the framework is based on proper scoring rules. Interestingly, ideas from cooperative game theory also come into play (even though mechanism design is usually based on noncooperative game theory). We propose several specific mechanisms based on concepts such as the Shapley value and VCG mechanisms.

It should immediately be noted that direct-revelation prediction markets, strictly interpreted, are probably not practical. This is because agents in such a market need to be able to directly reveal all their information that is pertinent to the prediction, including for example facts such as "I have interacted with some of the Democratic candidate's staff and they seem very motivated and inspired." It is hard to imagine a mechanism that can directly take such arbitrary natural language statements as input, determine a corresponding prediction in the form of a probability, and reward agents appropriately. Still, it seems worthwhile to study direct-revelation prediction markets, for at least the following reasons. A good theory of mechanism design for prediction markets should provide fundamental insight into the problem. It could serve as a natural starting point for the design of more practical markets: we could extend, simplify, or otherwise modify parts of the general theory when this seems appropriate for the setting at hand.[1] Also, towards the end of this paper, we will consider more practical designs based on the ideas in this paper.

## 2 Background

In this section, we review the necessary background in proper scoring rules and market scoring rules, mechanism design, and cooperative game theory.

### 2.1 Proper scoring rules & market scoring rules

Suppose our goal is to incentivize a single agent to truthfully report her subjective probability $p_E$ that an event $E$ will take place. We can do this by paying the agent some amount of money that depends both on the reported probability $\hat{p}_E$, and on whether the event actually occurs. If we let $x_E = 1$ if the event occurs, and $x_E = 0$ otherwise, then the agent receives a payment $s(\hat{p}_E, x_E)$. $s$ is said to be a *proper scoring rule* if the agent (uniquely) maximizes her (subjective) expected payoff by giving her true estimate of the probability—that is, for any $p \in [0, 1]$, $\{p\} = \arg\max_{\hat{p}} ps(\hat{p}, 1) + (1 - p)s(\hat{p}, 0)$. For simplicity, we will only consider settings with two outcomes—the event occurs, or it does not—but all of this is easily generalized to settings with more outcomes (for example, by running a separate proper scoring rule for each outcome).[2] Example proper scoring rules include the *quadratic* scoring rule, $1 - (x_E - \hat{p}_E)^2$ [1], and the *logarithmic* scoring rule, $x_E \log \hat{p}_E + (1 - x_E) \log(1 - \hat{p}_E)$ [8]. For the purpose of this paper, any proper scoring rule will do.

Can a proper scoring rule be used in a setting with multiple agents? One elegant way of doing this is to use a *market scoring rule* [10].[3] In a market scoring rule, there is a current estimate of the probability, $\hat{p}_E$. At any point in time, any agent can change the current probability to any $\hat{p}'_E$. In the end, this agent will be paid $s(\hat{p}'_E, x_E) - s(\hat{p}_E, x_E)$ for this change (which may be negative). This in some sense still gives the right incentive to the agent, because the agent cannot affect $s(\hat{p}_E, x_E)$. However, a market scoring rule still suffers from both of the problems discussed in the introduction: it can happen that not all the information is aggregated, and an agent may misreport in order to mislead other agents and take advantage later. In fact, both examples from the introduction are easily modified to the language of market scoring rules. One nice property of market scoring rules is that almost all the payment terms cancel out, so that the net payment made by the rule is $s(\hat{p}_E^f, x_E) - s(\hat{p}_E^0, x_E)$, where $\hat{p}_E^f$ is the final probability and $\hat{p}_E^0$ is the initial probability.

---

[1]Even in auction theory, which interacts more fluidly with mechanism design, the theoretical auction mechanisms from mechanism design (such as the generalized Vickrey auction) are generally not considered immediately practical, but they help establish a framework that is helpful in the design of more practical mechanisms, such as ascending auctions.

[2]This is not at all intended to give the impression that, in general, there is no reason to study prediction markets with more than two outcomes. Especially settings with a combinatorial outcome space lead to a variety of nontrivial and important research questions (*e.g.*, [7, 4, 3]).

[3]Other schemes for using proper scoring rules in a multiagent setting have been proposed [11, 12]; for the sake of brevity, we do not discuss them here.

## 2.2 Mechanism design

In a typical mechanism design setting, there are $n$ agents. Each of these agents has some private information, also known as a *signal*. Agent $i$'s signal is $\theta_i \in \Theta_i$, where $\Theta_i$ is the set of signals that $i$ might receive. The signal often represents the agent's personal preferences—for example, in an auction, her valuation for the item for sale—but the signal can also represent other types of private information—for example, the agent may know whether the item is authentic. In a *direct-revelation mechanism*, each agent $i$ reports some $\hat{\theta}_i \in \Theta_i$, not necessarily equal to her true signal. The mechanism is defined by a function $f : \Theta_1 \times \ldots \times \Theta_n \to O$, where $O$ is the set of all possible outcomes in the domain. The outcome $f(\hat{\theta}_1, \ldots, \hat{\theta}_n)$ can describe such things as the resulting allocation of resources, payments to be made by/to the agents, *etc*. Letting $\Theta = \Theta_1 \times \ldots \times \Theta_n$, each agent $i$ has a *utility function* $u_i : \Theta \times O$, where $u_i(\theta, o)$ gives the agent's utility when the agents' true signals are $\theta$ and the outcome chosen is $o$. (Often, the agent's utility depends only on her own signal $\theta_i$, but this is not necessarily the case. For example, even in an auction, one agent may know something about the item for sale that affects another agent's utility for the item.)

Of particular interest are mechanisms that incentivize agents to report their signals truthfully; such mechanisms are called *incentive compatible*. Let $\theta_{-i}$ denote the vector of all signals, with the exception of $i$'s signal. A mechanism is *ex-post incentive compatible* if for every $i$, $\theta_i, \hat{\theta}_i \in \Theta_i, \theta_{-i} \in \Theta_{-i}$, we have

$$u_i((\theta_i, \theta_{-i}), f(\theta_i, \theta_{-i})) \geq u_i((\theta_i, \theta_{-i}), f(\hat{\theta}_i, \theta_{-i}))$$

That is, each agent is best off reporting truthfully, regardless of the signals that the other agents receive, as long as the other agents report truthfully as well. A weaker notion, which requires us to have a prior distribution over the joint signal space $\Theta$, is *ex-interim incentive compatibility*. This requires that for every $i$, $\theta_i, \hat{\theta}_i \in \Theta_i$, we have

$$E_{\theta_{-i}|\theta_i}[u_i((\theta_i, \theta_{-i}), f(\theta_i, \theta_{-i}))] \geq E_{\theta_{-i}|\theta_i}[u_i((\theta_i, \theta_{-i}), f(\hat{\theta}_i, \theta_{-i}))]$$

That is, each agent is, *in expectation over the others' signals*, best off reporting truthfully, as long as the other agents report truthfully as well.

## 2.3 Cooperative game theory

A common object of study in cooperative game theory is a *characteristic function game*. Here, there is a set of agents $1, \ldots, n$, and a characteristic function $v : 2^{\{1,\ldots,n\}} \to \mathbb{R}$. For any subset (*coalition*) $C$ of the agents, $v(C)$ represents the value that that coalition can generate by working together. A key question is how the total value generated by the *grand coalition* of all agents—$v(\{1, \ldots, n\})$—should be distributed among the agents. One way to do so is to impose an ordering on the agents, represented by a permutation $\pi$ where $\pi(j)$ is the agent ranked $j$th in the ordering. Then, we give agent $i$, who is ranked in the $\pi^{-1}(i)$th position, her *marginal contribution*, which is $v(\{\pi(1), \pi(2), \ldots, \pi(\pi^{-1}(i))\}) - v(\{\pi(1), \pi(2), \ldots, \pi(\pi^{-1}(i) - 1)\})$ (where, of course, $\pi(\pi^{-1}(i)) = i$). Under the marginal contribution scheme, the value distribution in general depends heavily on the order $\pi$ chosen. For settings where there is no natural order on the agents, a reasonable approach is simply to average over *all* possible orders. Hence, agent $i$ receives $(1/n!) \sum_\pi v(\{\pi(1), \pi(2), \ldots, \pi(\pi^{-1}(i))\}) - v(\{\pi(1), \pi(2), \ldots, \pi(\pi^{-1}(i) - 1)\})$. This value distribution scheme is known as the *Shapley value* [15].

## 3 State-based model for prediction markets

We are now ready to introduce the framework that we will consider. We suppose that there is a set of *states* $S$ that the world can be in, as well as a common prior $P$ over these states. Our goal is to assess the probability of some event $E \subseteq S$, given the agents' information. Each agent $i$ has private information $S_i \subseteq S$, which consists of the states that are consistent with $i$'s information. That is, agent $i$ can rule out the states $S \setminus S_i$. In our framework, an agent is to report her full information $S_i$ directly, rather than just reporting a probability. (We assume that no new information enters the system over time and do not consider dynamic or iterative mechanisms.) Given this, we might conclude that the combined information of all the agents is $\bigcap_i S_i$, and given this we have a conditional probability for the event of $P(E | \bigcap_i S_i)$.[4] In fact, the notion that the conditional probability for the event is $P(E | \bigcap_i S_i)$ is not as straightforward as it may appear. Consider the following example:

**Example 1** *Consider a two-state example with $S = \{a, b\}$, where $E$ is true for $a$ and false for $b$ (so that we are predicting the probability that $a$ happens), and the prior is uniform. Suppose that there is only a single agent. If the true state is $a$, then with probability .5, the agent receives* the signal $\theta = \{a\}$ (she can rule out state $b$), and with probability $.5$, she receives the signal $\theta = \{a, b\}$ (she cannot rule out any state). If the true state is $b$, then with probability $1$ she receives the signal $\theta = \{a, b\}$. We note that the signal is always consistent with the true state, and, indeed, none of the states in a signal can be ruled out. However, we have $P(E|\theta = \{a, b\}) = P(E \wedge [\theta = \{a, b\}])/P(\theta = \{a, b\}) = (1/4)/(1/4 + 1/2) = 1/3$. In contrast, we have $P(E|\{a, b\}) = P(E) = 1/2$ (because $P(E|\{a, b\})$ is the probability that $E$ happens given that the true state is $a$ or $b$, which is always true).

---

[4]It is once again desirable to immediately emphasize the difficulties with turning such a framework into a practical design for a prediction market: we need a description of all possible states of the world (corresponding to all information that can possibly be revealed), and moreover we need a prior over these states.

Effectively, in Example 1, the signal contains information in addition to which states it rules out. The following definition considers models where this does not happen:

**Definition 1** *The model is* consistent *if, for every subset $C$ of the agents, for every combination $\theta_C$ of $|C|$ signals that the agents in $C$ can receive (where $\theta_i = S_i \subseteq S$), we have $P(E|\theta_C) = P(E|\bigcap_{i \in C} S_i)$.*

We now show that the inconsistency in Example 1 is due to the randomness of the signals.

**Definition 2** *We say that* signals are deterministic *if for every agent $i$, there is a partition $S_i^1, S_i^2, \ldots, S_i^{k_i}$ of $S$, so that if the true state is $s \in S_i^j$, then agent $i$ is guaranteed to get signal $S_i^j$.*

**Proposition 1** *If signals are deterministic, then the model is consistent.*

**Proof**: For every subset $C$ of the agents, for every combination $\theta_C$ of $|C|$ signals that the agents in $C$ can receive (where $\theta_i = S_i \subseteq S$), we have $P(E|\theta_C) = P(E|\theta_C \wedge \bigcap_{i \in C} S_i) = P(E|\bigcap_{i \in C} S_i)$, because $\theta_C$ happens if and only if $\bigcap_{i \in C} S_i$ happens. ∎

Now, a simple trick to make sure that signals are deterministic is to simply make them part of the state space. The next example illustrates this.

**Example 2** *Consider Example 1. We extend the state space to have three states: $a_1 = (a \wedge [\theta = \{a\}]), a_2 = (a \wedge [\theta = \{a, b\}]), b_1 = (b \wedge [\theta = \{a, b\}])$. These states happen with probabilities $1/4, 1/4, 1/2$, respectively. We have $E = \{a_1, a_2\}$. In this modified state space, the signal $\theta'$ that the agent receives is either $\{a_1\}$ or $\{a_2, b_1\}$, so signals are deterministic. Indeed, $P(E|\theta' = \{a_2, b_1\}) = P(E|\{a_2, b_1\}) = (1/4)/(1/4 + 1/2) = 1/3$.*

## 4 Some specific mechanisms

We now consider some specific mechanisms for rewarding the agents for the information that they contribute. To do so, we can use any fixed proper scoring rule $s$.

### 4.1 Rewarding agents individually

Presumably, the simplest approach is the following. Given the common prior $P$, each agent's reported individual information $\hat{S}_i$ leads to a probability estimate $P(E|\hat{S}_i)$, for which we can reward the agent with the proper scoring rule.

**Definition 3** *Under the* individual-rewarding information mechanism, *agent $i$ receives $s(P(E|\hat{S}_i), x_E)$.*

**Proposition 2** *If the model is consistent, then the individual-rewarding information mechanism is ex-interim incentive compatible.*

**Proof**: By consistency, we have $P(E|\theta_i) = P(E|S_i)$. Because $s$ is a proper scoring rule, the agent maximizes her expected utility by ensuring this probability is entered into the scoring rule, which can be done by reporting truthfully ($\hat{S}_i = S_i$). ∎

It should be observed that reporting truthfully is not necessarily the *uniquely* optimal action, because multiple reports $\hat{S}_i$ may each lead to the same probability $P(E|\hat{S}_i)$. It is also not *ex-post* incentive compatible, because given the other agents' signals, the agent would in general prefer to report that information as well (i.e., report $\bigcap_j S_j$), to get a better probability estimate.

There are additional downsides of this mechanism. For one, in principle, this may result in payments to agents who do not contribute any information (that is, agents who report $S$): $s(P(E|S), x_E)$ may not be zero. This is easy to fix, by paying an agent who reports $\hat{S}_i$ an amount of $s(P(E|\hat{S}_i), x_E) - s(P(E|S), x_E)$ instead, so that an agent who reports no information receives nothing. Equivalently, we can modify (shift) the scoring rule to $s'(\hat{p}_E, x_E) = s(\hat{p}_E, x_E) - s(P(E|S), x_E)$. Hence, we can assume without loss of generality that $s(P(E|S), x_E) = 0$ (no payments for no information).

However, there are other oddities. For example, suppose that agents 1 and 2 each have the same information, $S' \subseteq S$, and agent 3 has different information, $S'' \subseteq S$. Furthermore, suppose that $P(E|S') = P(E|S'') = 0.7$, and $P(E|S' \cap S'') = 0.9$; and suppose that the event indeed happens ($x_E = 1$). Then, each of the three agents receives the same payment $s(0.7, 1)$. However, it intuitively seems to make more sense to reward the third agent more, because her information $S''$ was more unique (and no less relevant). In effect, we are paying twice for the same information, $S'$. This can also lead to excessive payments: if many agents all report $S'$, we still have to pay each of them $s(0.7, 1)$, even though none of them (except for the first one) contribute any new information.

### 4.2 Rewarding based on marginal information

A different approach is to pay an agent only for the *marginal* information that that agent reports.

**Definition 4** *If the agents are ordered $1, \ldots, n$, then under the* marginal information mechanism*, agent $i$ receives* $s(P(E|\bigcap_{j=1,\ldots,i} \hat{S}_j), x_E) - s(P(E|\bigcap_{j=1,\ldots,i-1} \hat{S}_j), x_E)$.

This is almost the same idea as using a market scoring rule, in which there is a market estimate $\hat{p}_E$ of the probability of the event, and if an agent shifts the market probability from $\hat{p}_E$ to $\hat{p}'_E$, she receives $s(\hat{p}'_E, x_E) - s(\hat{p}_E, x_E)$. If each agent can only move the market probability once, and they do so in the order $1, \ldots, n$, then, if $\hat{p}^i_E$ is the market probability after $i$'s move, agent $i$ receives $s(\hat{p}^i_E, x_E) - s(\hat{p}^{i-1}_E, x_E)$. We note that agent $i$ can observe the sequence $\hat{p}^0_E, \ldots, \hat{p}^{i-1}_E$ before making her move. If we apply this market scoring rule to our model, *if* the observed sequence is sufficient for each agent $i$ to infer the earlier agents' information $\bigcap_{j=1,\ldots,i-1} S_j$, then she will move the probability to $P(E|\bigcap_{j=1,\ldots,i} S_j)$. So, in this case, the market scoring rule and the marginal information mechanism described above will produce the same result (given truthful behavior). However, the sequence of market probabilities in general is not sufficient to infer the previous agents' information (*e.g.*, consider the first example in the introduction), and as a result, the final probability $\hat{p}^n_E$ sometimes does not reflect all the information available to the agents.

In contrast, under the marginal information mechanism, the information is revealed directly (rather than indirectly in the form of a probability), so that the final probability $P(E|\bigcap_{j=1,\ldots,n} \hat{S}_j)$ does reflect all the information available to the agents (given truthful behavior, that is, $\hat{S}_j = S_j$). We also note that in the marginal information mechanism, the agents themselves do not need to reason about how the probability should be updated; rather, they just report their information (simultaneously), and the mechanism automatically figures out the probabilities based on the common prior. Of course, the market scoring rule is easier to apply in practice because it does not require a common prior or a model of the information that agents might report, but we will discuss practical issues later. The marginal information approach shares some of the nice properties of a market scoring rule: just as the market scoring rule in total pays exactly $s(\hat{p}^n_E, x_E) - s(\hat{p}^0_E, x_E)$, the marginal information mechanism in total pays exactly $s(P(E|\bigcap_{j=1,\ldots,n} \hat{S}_j), x_E) - s(P(E|S), x_E)$. So, both mechanisms pay exactly for the total information that they receive.

**Proposition 3** *If the model is consistent, then the marginal information mechanism is ex-interim incentive compatible.*

**Proof**: We will prove that it is optimal for agent $i$ to report truthfully even if she knows $S_1, \ldots, S_{i-1}$ (but none of $S_{i+1}, \ldots, S_n$), regardless of what $S_1, \ldots, S_{i-1}$ are. From this, it follows that it is also optimal to report truthfully if she does not know these. Since the other agents are assumed to tell the truth, we have $S_j = \hat{S}_j$ for $j \neq i$. Agent $i$ will receive $s(P(E|\bigcap_{j=1,\ldots,i} \hat{S}_j), x_E) - s(P(E|\bigcap_{j=1,\ldots,i-1} \hat{S}_j), x_E)$. Because she cannot affect $s(P(E|\bigcap_{j=1,\ldots,i-1} \hat{S}_j), x_E)$, her goal is to maximize $s(P(E|\bigcap_{j=1,\ldots,i} \hat{S}_j), x_E)$. By consistency, we have $P(E|\theta_1 = S_1, \ldots, \theta_i = S_i) = P(E|\bigcap_{j=1,\ldots,i} S_j)$. Because $s$ is a proper scoring rule, agent $i$ maximizes her expected utility by ensuring this probability is entered into the scoring rule. This can be done by reporting truthfully ($\hat{S}_i = S_i$), because $\hat{S}_j = S_j$ for $j < i$. ∎

### 4.3 Shapley value information mechanism

As in the case of characteristic function games, there may not be a natural order on the agents to use in the marginal information mechanism; indeed, this seems unnatural for a direct-revelation mechanism. We can employ the same solution as in characteristic function games: simply take the average over all possible orders of the agents.

**Definition 5** *Under the* Shapley value information mechanism*, agent $i$ receives the average, taken over all orders of the agents, of what she would have received under the marginal information mechanism.*

Because the Shapley value information mechanism is simply the average of all marginal information mechanisms, it also pays exactly $s(P(E|\bigcap_{j=1,\ldots,n} \hat{S}_j), x_E) - s(P(E|S), x_E)$ in the end.

**Proposition 4** *If the model is consistent, the Shapley value information mechanism is ex-interim incentive compatible.*

**Proof**: This follows immediately from the fact that the Shapley value information mechanism is an average of marginal information mechanisms, which are ex-interim incentive compatible by Proposition 3. ∎

### 4.4 Conversion to characteristic function games

The marginal information mechanism, and (as a result) the Shapley value information mechanism, are ways of distributing the total value $s(P(E|\bigcap_{i=1,\ldots,n} \hat{S}_i), x_E) - s(P(E|S), x_E)$ across the agents. We can describe this directly as a characteristic function game, as follows. The value that would be generated by a coalition $C \subseteq \{1, \ldots, n\}$ in the absence of the other agents is $s(P(E|\bigcap_{j \in C} \hat{S}_j), x_E) - s(P(E|S), x_E)$. With this insight, we can use *any* solution concept from cooperative game theory, not just the marginal contribution mechanism or the Shapley value.

We may wonder if the characteristic function games resulting from this conversion have some special structure. For one, the values that coalitions can have are limited by the range of the proper scoring rule used. Apart from the proper scoring rule used, what matters is the probability of the event given the information of a subset of agents. Let us consider the *pre-characteristic function* $\pi : 2^{\{1,\ldots,n\}} \to [0,1]$, where $\pi(C) = P(E|\bigcap_{j \in C} \hat{S}_j)$. The following theorem shows that this function can take any form, as long as probabilities are not equal to 0 or 1.

**Theorem 1** *For any function* $\psi : 2^{\{1,\ldots,n\}} \to (0,1)$, *there exists an instance such that for all* $C$, $\pi(C) = P(E|\bigcap_{j \in C} \hat{S}_j) = \psi(C)$.

**Proof**: Suppose that each agent $i$ receives a signal $x_i \in \{0,1\}$. We will construct a prior such that, if all agents receive a signal of 1, then the pre-characteristic function $\pi(C)$ coincides with $\psi(C)$. To do so, we construct a potential function $\chi : \{0,1\}^n \to \mathbb{R}^{\geq 0}$, so that for any vector of signals $\theta \in \{0,1\}^n$, we have $P(\theta) = \chi(\theta)/\sum_{\theta' \in \{0,1\}^n} \chi(\theta')$. We also construct a function $\chi_E : \{0,1\}^n \to \mathbb{R}^{\geq 0}$, so that for any vector of signals $\theta \in \{0,1\}^n$, we have $P(E|\theta) = \chi_E(\theta)/\chi(\theta)$.

We now show how $\chi$ and $\chi_E$ can be chosen so that we obtain the desired property that for all $C \subseteq \{0,1\}^n$, we have $P(E|(\forall i \in C) \, \theta_i = 1) = \psi(C)$ for all $C$. For any $\theta \in \{0,1\}^n$, let $c(\theta)$ be the number of 1s in $\theta$. We will start by setting $\chi$ and $\chi_E$ for the $\theta$ with large $c(\theta)$, and work our way down to $\theta$ with smaller $c(\theta)$. Say that $\theta < \theta'$ if $\theta'$ has a 1 in every place where $\theta$ has a 1. Let $\theta^C \in \{0,1\}^n$ be the vector of signals where every member of $C$ has a 1, and every other agent has a 0. We have that $P(E|(\forall i \in C) \, \theta_i = 1) = \frac{\sum_{\theta \geq \theta^C} \chi_E(\theta)}{\sum_{\theta \geq \theta^C} \chi(\theta)}$. Now, suppose that we have set $\chi$ and $\chi_E$ for all $\theta$ with $c(\theta) \geq k$, in such a way that for every $C$ with $|C| \geq k$, $P(E|(\forall i \in C) \, \theta_i = 1) = \psi(C)$. We will show that we can set $\chi$ and $\chi_E$ for the $\theta$ with $c(\theta) = k-1$ in such a way that for every $C$ with $|C| = k-1$, $P(E|(\forall i \in C) \, \theta_i = 1) = \psi(C)$. This is because for any such $C$, we have $P(E|(\forall i \in C) \, \theta_i = 1) = \frac{\sum_{\theta \geq \theta^C} \chi_E(\theta)}{\sum_{\theta \geq \theta^C} \chi(\theta)}$, where all the terms in the denominator and the numerator have already been determined, with the exception of $\chi(\theta^c)$ in the denominator and $\chi_E(\theta^c)$ in the numerator. Because $0 < \psi(C) < 1$, these two can be set so that $P(E|(\forall i \in C) \, \theta_i = 1) = \psi(C)$. ∎

This tells us that the characteristic functions in our model have very little structure.

### 4.5 Rewarding based on the group's information

None of the mechanisms proposed so far are ex-post incentive compatible, because if an agent knew the other agents' private information, she would generally prefer to report that information as well. We now consider a simple mechanism that is ex-post incentive compatible.

**Definition 6** *Under the* group-rewarding information mechanism, *every agent $i$ receives the same amount,* $s(P(E|\bigcap_{j=1,\ldots,n} \hat{S}_j), x_E)$.

**Proposition 5** *If the model is consistent, then the group-rewarding information mechanism is ex-post incentive compatible.*

**Proof**: We consider the case where agent $i$ knows all the other agents' signals (as well as her own). By consistency, we have $P(E|\theta_1 = S_1, \ldots, \theta_n = S_n) = P(E|\bigcap_{j=1,\ldots,n} S_i)$. Because $s$ is a proper scoring rule, agent $i$ maximizes her expected utility by ensuring this probability is entered into the scoring rule. This can be done by reporting truthfully ($\hat{S}_i = S_i$), because by assumption, the other agents report truthfully, so that $\hat{S}_j = S_j$ for $j \neq i$. ∎

The group-rewarding mechanism has the awkward property that an agent who reports no information is paid the same as an agent who reports a lot of information. The next subsection shows how to avoid this.

### 4.6 Pivotal information mechanism

We can modify the group-rewarding information mechanism as follows, to obtain a mechanism like the *pivotal* (or *Clarke*) mechanism [5, 9].

**Definition 7** *Under the* pivotal information mechanism, *agent $i$ receives* $s(P(E|\bigcap_{j=1,\ldots,n} \hat{S}_j), x_E) - s(P(E|\bigcap_{j \neq i} \hat{S}_j), x_E)$.

**Proposition 6** *If the model is consistent, then the pivotal information mechanism is ex-post incentive compatible.*

**Proof**: Agent $i$ cannot affect the term $s(P(E|\bigcap_{j \neq i} \hat{S}_j), x_E)$, so her incentives are the same as under the group-rewarding information mechanism. ∎

From the proof, it is clear that we can add *any* term that does not depend on $i$'s report to the payment in the group-rewarding information mechanism, and the ex-post incentive compatibility property will be maintained. This is analogous to the class of *Groves* mechanisms [9]. We now give an axiomatization of the pivotal information mechanism.

**Definition 8** *A mechanism satisfies the* individual agent *property if, for every $i$, we have the following: given that the other agents report no information ($\hat{S}_j = S$ for all $j \neq i$), agent $i$ receives* $s(P(E|S_i), x_E) - s(P(E|S), x_E)$.

**Definition 9** *A mechanism satisfies the* strong decomposition *property if, for every $i$, for every $S', S'', S''' \subseteq S$, we*

have the following: if agent $i$ reports $\hat{S}_i = S' \cap S''$, and the other agents report $\bigcap_{j \neq i} \hat{S}_j = S'''$, then the reward that $i$ gets is equal to the reward that she would get if she reported $S''$ and the other agents reported $S'''$, plus the reward that she would get if she reported $S'$ and the other agents reported $S'' \cap S'''$. It satisfies the weak decomposition property *if the above holds when $S''' = S$.*

**Definition 10** *A mechanism satisfies the* relative dummy *property if, for every $i$, given that $i$ reports no information that is not also reported by the other agents (that is, $\bigcap_{j \neq i} \hat{S}_j \subseteq \hat{S}_i$), agent $i$ receives $0$.*

**Lemma 1** *The weak decomposition property implies the relative dummy property.*

**Proof**: Let $S' = \hat{S}_i$ and $S'' = \bigcap_{j \neq i} \hat{S}_j$, where $S'' \subseteq S'$. Then, given weak decomposition, the reward an agent gets for reporting $S' \cap S'' = S''$ when nobody else reports any information is equal to the reward the agent gets for reporting $S''$ when nobody else reports any information, plus the reward the agent gets for reporting $S'$ when the other agents report $S''$. Because the first two terms are the same, the third must be zero. Hence, the relative dummy property must be satisfied. ∎

**Theorem 2** *The pivotal information mechanism satisfies the individual agent, strong decomposition, and relative dummy properties. No other mechanism satisfies the individual agent and weak decomposition properties.*

**Proof**: First we show that the pivotal information mechanism satisfies these properties. If the other agents report no information, then $\bigcap_{j \neq i} \hat{S}_j = S$, so that agent $i$'s payment is $s(P(E|\bigcap_{j=1,\ldots,n} \hat{S}_j), x_E) - s(P(E|\bigcap_{j \neq i} \hat{S}_j), x_E) = s(P(E|S_i), x_E) - s(P(E|S), x_E)$. Therefore, the pivotal information mechanism satisfies the individual agent property. If agent $i$ reports $\hat{S}_i = S' \cap S''$, and the other agents report $\bigcap_{j \neq i} \hat{S}_j = S'''$, then agent $i$'s reward is $s(P(E|S' \cap S'' \cap S'''), x_E) - s(P(E|S'''), x_E)$, which can be rewritten as $[s(P(E|S' \cap S'' \cap S'''), x_E) - s(P(E|S'' \cap S'''), x_E)] + [s(P(E|S'' \cap S'''), x_E) - s(P(E|S'''), x_E)]$. The second part in brackets is the reward that she would get if she reported $S''$ and the other agents reported $S'''$, and the first part in brackets is the reward that she would get if she reported $S'$ and the other agents reported $S'' \cap S'''$. Therefore, the pivotal information mechanism satisfies the strong decomposition property. The strong decomposition property implies the weak decomposition property, which in turn, by Lemma 1, implies the relative dummy property.

Conversely, consider a mechanism that satisfies the individual agent and weak decomposition properties. Let $S' = \hat{S}_i$ and let $S'' = \bigcap_{j \neq i} \hat{S}_j$. By weak decomposition, we have that the reward for reporting $S' \cap S''$ when nobody else reports any information is equal to the reward for reporting $S''$ when nobody else reports any information, plus the reward for reporting $S'$ when the other agents report $S''$. Equivalently, the reward for reporting $S'$ when the other agents report $S''$ is equal to the reward for reporting $S' \cap S''$ when nobody else reports any information, minus the reward for reporting $S''$ when nobody else reports any information. By the individual agent property, it follows that the reward for reporting $S'$ when the other agents report $S''$ is equal to $[s(P(E|S' \cap S''), x_E) - s(P(E|S), x_E)] - [s(P(E|S''), x_E) - s(P(E|S), x_E)] = s(P(E|S' \cap S''), x_E) - s(P(E|S''), x_E) = s(P(E|\bigcap_{j=1,\ldots,n} \hat{S}_j), x_E) - s(P(E|\bigcap_{j \neq i} \hat{S}_j), x_E)$. So it must be the pivotal information mechanism. ∎

## 5 Creating a practical design with information and probability agents

As already mentioned, the model described in the previous sections, in spite of its theoretical advantages, does not directly lead to a natural design for a prediction market, because using it directly requires a model of all the information that can be reported—that is, a model of all the possible states of the world—and a prior distribution over these states. In regard to the latter, the purpose of a typical prediction market is precisely to get such probabilistic information from the participating agents!

To address this, let us suppose that we can divide the agents participating in a prediction market into two categories, *information agents* and *probability agents*. Information agents should be thought of as people that have relevant information about the event we are trying to predict—for example, someone who is friendly with some of the Democratic candidate's staff—but are not necessarily able to convert such information into a probability for the event. On the other side, probability agents do not necessarily have any information about the event, but they are able to convert any information that is given to them into a probability. For example, we can think of a person who has worked on a large number of presidential campaigns but is now retired and no longer has any direct access to what is happening in the campaign; nevertheless, if this person is told that the candidate's staff is very motivated and inspired, she can assess how this information changes the probability that the candidate will win the election, based on her many years of experience. This feels like a natural division: the people who have information relevant to an event are generally not the same as the people who are comfortable turning such information into an exact number reflecting the probability of the event. There may be agents that can do both; this is not necessarily a problem, but in the remainder we will think of these two groups of agents as separate.

This separation between information agents and probability agents allows us to run a version of the mechanisms described earlier in the paper, as follows. In order to run those mechanisms, all that is needed is that, given a collection of

```
Initialize p̂_E to some value
P̂_0(E) ← p̂_E
for j = 1 to n_2 {
  Probability agent j moves p̂_E to a new value
  P̂_j(E) ← p̂_E
}
P̂(E) ← P̂_{n_2}(E)
for i = 1 to n_1 {
  Information agent i reports information I_i (in natural language)
  P̂_0(E|I_1, . . . , I_i) ← p̂_E
  for j = 1 to n_2 {
    Probability agent j moves p̂_E to a new value
    P̂_j(E|I_1, . . . , I_i) ← p̂_E
  }
  P̂(E|I_1, . . . , I_i) ← P̂_{n_2}(E|I_1, . . . , I_i)
}
The event is realized to obtain x_E
for i = 1 to n_1
  Reward information agent i with s_1(P̂(E|I_1, . . . , I_i), x_E) −
  s_1(P̂(E|I_1, . . . , I_{i−1}), x_E)
for j = 1 to n_2
  Reward probability agent j with ∑_{i=0}^{n_1} [s_2(P̂_j(E|I_1, . . . , I_i), x_E) −
  s_2(P̂_{j−1}(E|I_1, . . . , I_i), x_E)]
```

Figure 1: A prediction market based on the marginal information mechanism and a market scoring rule.

information (that is, given a restricted set of states), we can compute the conditional probability of the event happening. Now, we can use the probability agents to give us estimates of these conditional probabilities, using a standard prediction market. That is, the information agents are playing in the information-based mechanism, and the probability agents are effectively used as a subroutine in this mechanism, to compute the conditional probabilities that running this information mechanism requires.

To illustrate the general principle, Figure 1 gives a complete design that combines the marginal information mechanism from earlier in the paper for the information agents (using scoring rule $s_1$) with a market scoring rule for the probability agents (using scoring rule $s_2$). The information agents report their information one at a time; after one of them has revealed her information, each of the probability agents gets a chance to move the market estimate of the probability. Let $n_1$ be the number of information agents and $n_2$ the number of probability agents. $\hat{p}_E$ is always the current market assessment of the probability of the event, which the probability agents can move around after obtaining new information. $I_i$ is the information that information agent $i$ reports (which we previously represented as $\hat{S}_i$, but it is perhaps less natural to think of information as a subset of states at this point). $\hat{P}_j(E|I_1, \ldots, I_i)$ is the market probability after information agents 1 through $i$ have reported their information, and, after $i$'s report, probability agents 1 through $j$ have updated the probability. $\hat{P}(E|I_1, \ldots, I_i) = \hat{P}_{n_2}(E|I_1, \ldots, I_i)$ is the market probability after all probability agents have updated the probability, after seeing 1 through $i$'s information. We omit discussion of this design's properties for the sake of space.

## 6 Future research

There are many directions for future research. On the theoretical side, new information mechanisms can be designed in this framework, and the properties of the mechanisms can be studied further. Many phenomena that occur in other mechanism design domains (such as auctions) have analogues in this domain. On the practical side, it would be interesting to put a design such as Figure 1 into practice.

## Acknowledgments

I thank Yiling Chen, David Pennock, Daniel Reeves, and the UAI reviewers for helpful discussions and detailed feedback. This work is supported by NSF IIS-0812113, the Sloan Foundation, and a Yahoo! Faculty Research Grant.